\documentclass[portrait]{emulateapj} 
\usepackage{natbib}

\shorttitle{SAGE-Spec: Overview and initial results}
\shortauthors{F.~Kemper et al.}

\begin{document}

\title{The SAGE-Spec Spitzer Legacy program: The life-cycle of dust
and gas in the Large Magellanic Cloud}
\author{F.~Kemper\altaffilmark{1}, Paul M.~Woods\altaffilmark{1},
V.~Antoniou\altaffilmark{2}, J.-P.~Bernard\altaffilmark{3},
R.~D.~Blum\altaffilmark{4}, M.~L.~Boyer\altaffilmark{5},
J.~Chan\altaffilmark{6}, C.-H.~R.~Chen\altaffilmark{7},
M.~Cohen\altaffilmark{8}, C.~Dijkstra\altaffilmark{9},
C.~Engelbracht\altaffilmark{10}, M.~Galametz\altaffilmark{11},
F.~Galliano\altaffilmark{11}, C.~Gielen\altaffilmark{12}, Karl
D.~Gordon\altaffilmark{5}, V.~Gorjian\altaffilmark{13},
J.~Harris\altaffilmark{10}, S.~Hony\altaffilmark{11},
J.~L.~Hora\altaffilmark{14}, R.~Indebetouw\altaffilmark{7,15},
O.~Jones\altaffilmark{1}, A.~Kawamura\altaffilmark{16},
E.~Lagadec\altaffilmark{17,1}, B.~Lawton\altaffilmark{5},
J.~M.~Leisenring\altaffilmark{7}, S.~C.~Madden\altaffilmark{11},
M.~Marengo\altaffilmark{2,14}, M.~Matsuura\altaffilmark{18,19},
I.~McDonald\altaffilmark{1}, C.~McGuire\altaffilmark{1},
M.~Meixner\altaffilmark{5}, A.~J.~Mulia\altaffilmark{6},
B.~O'Halloran\altaffilmark{20}, J.~M.~Oliveira\altaffilmark{21},
R.~Paladini\altaffilmark{22}, D.~Paradis\altaffilmark{22},
W.~T.~Reach\altaffilmark{22}, D.~Rubin\altaffilmark{11},
K.~Sandstrom\altaffilmark{8,23}, B.~A.~Sargent\altaffilmark{5},
M.~Sewilo\altaffilmark{5}, B.~Shiao\altaffilmark{5},
G.~C.~Sloan\altaffilmark{24}, A.~K.~Speck\altaffilmark{6},
S.~Srinivasan\altaffilmark{25,26}, R.~Szczerba\altaffilmark{27},
A.~G.~G.~M.~Tielens\altaffilmark{28}, E.~van Aarle\altaffilmark{12},
S.~D.~Van Dyk\altaffilmark{22}, J.~Th.~van Loon\altaffilmark{21},
H.~Van Winckel\altaffilmark{12}, Uma P.~Vijh\altaffilmark{29},
K.~Volk\altaffilmark{5}, B.~A.~Whitney\altaffilmark{30},
A.~N.~Wilkins\altaffilmark{24}, A.~A.~Zijlstra\altaffilmark{1}}
\altaffiltext{1}{Jodrell Bank Centre for Astrophysics, Alan Turing
Building, School of Physics and Astronomy, The University of
Manchester, Oxford Road, Manchester, M13 9PL, UK}
\altaffiltext{2}{Department of Physics and Astronomy, Iowa State
University, Ames, IA 50011} \altaffiltext{3}{Centre d'\'Etude Spatiale
des Rayonnements, 9 Av.~du Colonel Roche, BP 44346, 31028 Toulouse
cedex 4, France} \altaffiltext{4}{NOAO, 950 North Cherry Avenue,
Tucson, AZ 85719} \altaffiltext{5}{Space Telescope Science Institute,
3700 San Martin Drive, Baltimore, MD 21218} \altaffiltext{6}{Physics
\& Astronomy Department, University of Missouri, Columbia, MO 65211}
\altaffiltext{7}{Department of Astronomy, University of Virginia,
P.O. Box 400325, Charlottesville, VA 22904} \altaffiltext{8}{Radio
Astronomy Laboratory, University of California at Berkeley, 601
Campbell Hall, Berkeley, CA 94720-3411}
\altaffiltext{9}{Passiebloemweg 31, 1338 TT Almere, The Netherlands}
\altaffiltext{10}{Steward Observatory, University of Arizona, 933
North Cherry Avenue, Tucson, AZ 85721} \altaffiltext{11}{Laboratoire
AIM, CEA/DSM - CNRS - Universit{\'e} Paris Diderot DAPNIA/Service
d'Astrophysique B\^{a}t.~709, CEA-Saclay F-91191 Gif-sur-Yvette
C{\'e}dex, France} \altaffiltext{12}{Instituut Voor Sterrenkunde,
KULeuven, Celestijnenlaan 200D, 3001 Leuven (Heverlee), Belgium}
\altaffiltext{13}{JPL/Caltech, MS 169-506, 4800 Oak grove Dr.,
Pasadena, CA 91109} \altaffiltext{14}{Harvard-Smithsonian Center for
Astrophysics, 60 Garden Street, MS 65, Cambridge, MA 02138-1516}
\altaffiltext{15}{National Radio Astronomy Observatory, 520 Edgemont
Road, Charlottesville, VA 22903} \altaffiltext{16}{Department of
Astrophysics, Nagoya University, Chikusa-Ku, Nagoya 464-01, Japan}
\altaffiltext{17}{ESO Headquarters Garching,
Karl-Schwarzschild-Str.~2, D-85748 Garching bei Muenchen, Germany}
\altaffiltext{18}{Institute of Origins, Department of Physics and
Astronomy, University College London, Gower Street, London WC1E 6BT,
UK } \altaffiltext{19}{Institute of Origins, Mullard Space Science
Laboratory, University College London, Holmbury St. Mary, Dorking,
Surrey RH5 6NT, UK } \altaffiltext{20}{Astrophysics Group, Imperial
College London, Blackett Laboratory, Prince Consort Road, London, SW7
2AZ,UK} \altaffiltext{21}{School of Physical \& Geographical Sciences,
Lennard-Jones Laboratories, Keele University, Staffordshire ST5 5BG,
UK} \altaffiltext{22}{Spitzer Science Center, California Institute of
Technology, MS 220-6, Pasadena, CA 91125}
\altaffiltext{23}{Max-Planck-Institut f\"ur Astronomie, D-69117
Heidelberg, Germany} \altaffiltext{24}{Department of Astronomy,
Cornell University, Ithaca, NY 14853-6801}
\altaffiltext{25}{Department of Physics and Astronomy, Johns Hopkins
University, Homewood Campus, Baltimore, MD 21218}
\altaffiltext{26}{Institut d'Astrophysique de Paris, 98 bis, Boulevard
Arago, Paris 75014, France} \altaffiltext{27}{N.~Copernicus
Astronomical Center, Rabianska 8, 87-100 Torun, Poland }
\altaffiltext{28}{Leiden Observatory, P.O. Box 9513, NL-2300 RA
Leiden, The Netherlands } \altaffiltext{29}{Ritter Astrophysical
Research Center, University of Toledo, Toledo OH 43606}
\altaffiltext{30}{Space Science Institute, 4750 Walnut Street, Suite
205, Boulder, CO 80301}

\begin{abstract} The \emph{SAGE-Spec} \emph{Spitzer} Legacy program is a
spectroscopic follow-up to the \emph{SAGE-LMC} photometric survey of
the Large Magellanic Cloud carried out with the \emph{Spitzer Space
Telescope}. We present an overview of SAGE-Spec and some of its first
results. The \emph{SAGE-Spec} program aims to study the life cycle of
gas and dust in the Large Magellanic Cloud, and to provide information
essential to the classification of the point sources observed in the
earlier \emph{SAGE-LMC} photometric survey. We acquired 224.6 hours of
observations using the \emph{InfraRed Spectrograph} and the SED mode of
the \emph{Multiband Imaging Photometer for Spitzer}. The
\emph{SAGE-Spec} data, along with archival \emph{Spitzer} spectroscopy of
objects in the Large Magellanic Cloud, are reduced and delivered to
the community. We discuss the observing strategy, the specific data
reduction pipelines applied and the dissemination of data products to
the scientific community. Initial science results include the first
detection of an extragalactic ``21 $\mu$m'' feature towards an evolved
star and elucidation of the nature of disks around RV\,Tauri stars in
the Large Magellanic Cloud.  Towards some young stars, ice features
are observed in absorption.  We also serendipitously observed a
background quasar, at a redshift of $z\approx 0.14$, which appears to
be host-less.

\end{abstract}


\section{Introduction}

A photometric survey in the infrared of the Large Magellanic Cloud
(LMC) was performed by the \emph{Spitzer} Legacy Program
\emph{Surveying the Agents of Galaxy Evolution}
\citep[SAGE-LMC;][]{2006AJ....132.2268M}, which charts the budget of
gas and dust contributing to the cycle of star formation and stellar
death in the Magellanic Clouds. Here we discuss \emph{SAGE-Spec}, a
spectroscopic follow-up to \emph{SAGE-LMC}, which is also a \emph{Spitzer}
Legacy Program. For \emph{SAGE-Spec} we observed a variety of
circumstellar and interstellar environments with the \emph{Infrared
Spectrograph} \citep[IRS;][]{2004ApJS..154...18H} aboard \emph{Spitzer}
\citep{2004ApJS..154....1W}, as well as the SED mode available on the
\emph{Multiband Imaging Photometer for Spitzer}
\citep[MIPS;][]{2004ApJS..154...25R}.  The \emph{SAGE-Spec} dataset is
exceptionally suited to address the following issues: First, it allows
us to trace the lifecycle of dust and molecular gas on its journey
through the galaxy, from dust production sites (AGB stars, red
supergiants, post-AGB-objects, planetary nebulae), to the ISM (atomic
and molecular clouds) to star forming regions (\ion{H}{2} regions,
young stellar objects); and, second, it allows us to develop a
photometric color-color and color-magnitude classification scheme to
increase the legacy of the larger \emph{SAGE-LMC} database.  In
addition, a large number of smaller astrophysical questions can be
addressed using the data set provided here, and a rich harvest in
scientific results is expected.

This paper gives an overview of the \emph{SAGE-Spec} Legacy program.
We outline the observing strategy, describe the data reduction process
and discuss the data products that are currently publicly available to
the astronomical community \citep{SAGESpec-delivery2}, or will become
publicly available in the near future.  Existing surveys targeting gas
and dust in the LMC are discussed in Sect.~\ref{sec:surveys}. This
section also includes a description of existing publications of
infrared spectroscopy on LMC targets.  The observing strategy of the
\emph{SAGE-Spec} project is discussed in Sect.~\ref{sec:observations},
along with a description of the data reduction.  This paper finishes
with some first scientific results of the \emph{SAGE-Spec} project
(Sect.~\ref{sec:results}), and conclusions and an outlook to the
future (Sect.~\ref{sec:conclusions}).

\section{Surveys: The gas and dust in the LMC}
\label{sec:surveys}

In order to study the life-cycle of dust on a galactic scale, the LMC
provides a good compromise between distance and size. It is found at a
distance of $\sim$50 kpc \citep{F_99_cepheids}, and as an additional
benefit has a favorable viewing angle
\citep[35$^{\mathrm{o}}$,][]{VC_01_viewingangle} resulting in low
column densities, and typically just a single interstellar cloud,
along each line-of-sight. It is possible to observe individual objects
in the LMC due to its vicinity, while at the same time the outside
viewpoint that we have enables us to obtain a global view of the LMC
through surveys like this.

With $Z\approx 0.3-0.5\,Z_{\odot}$, the metallicity of the LMC is
sub-solar \citep{W_97_MgClds}.  As a consequence the dust-to-gas mass
ratio is $\sim$2--4 times lower than that in the Solar neighborhood
\citep{GCM_03_extinction}, permitting easier penetration of UV
radiation to affect physical processes in the ISM and star formation.
Indeed, modeling of photon-dominated regions shows that molecular
clouds in the LMC will be larger and less dense than clouds in the
Milky Way \citep{1998ApJ...498..735P}. The shape of the UV
interstellar extinction curve appears to be independent of metallicity,
thus constraining the differences between Galactic and LMC grain
properties, although large variations exist between environments
within the LMC \citep{1999ApJ...515..128M}.

\subsection{Surveys of the LMC}

Previous infrared surveys of the stellar content of the LMC, performed
with, for instance, MSX \citep{EVD_01_LMC} and DENIS
\citep{CVL_00_redgiantbranch}, are limited to only the brightest
sources.  Exploration of the stellar content of the LMC was therefore
skewed to the tip of the AGB and some bright supergiants. However,
with its sensitive arrays, \emph{Spitzer} has allowed for a full
census of all objects brighter than $\sim 15^{\mathrm{th}}$ magnitude
in the 8.0 $\mu$m band \citep{2006AJ....132.2268M}.  \emph{SAGE-LMC}
encompasses a field of $7^{\mathrm{o}} \times 7^{\mathrm{o}}$ covering
the majority of the LMC, observed using all bands of the
\emph{InfraRed Array Camera} \citep[IRAC;][]{2004ApJS..154...10F} and
MIPS.  The IRAC and MIPS point source catalog contains about 6 million
sources, and has been made available to the community\footnote{The
\emph{SAGE-LMC} point source catalog can be accessed on {\tt
http://irsa.ipac.caltech.edu/applications/Gator/}}
\citep{2006AJ....132.2268M}. The post- and pre-Main Sequence
populations uncovered by \emph{SAGE-LMC} are discussed by
e.g.~\citet{BMO_06_evolved} and \citet{2008AJ....136...18W}.

The extended emission component of the \emph{SAGE-LMC} survey is
discussed in more detail by \citet{2008AJ....136..919B}, who include
references to additional LMC surveys in molecular and atomic emission,
e.g.~H$\alpha$ \citep{2001PASP..113.1326G}, H{\sc i}
\citep{2003ApJS..148..473K,SKC_03_HI}, and CO
\citep{2008ApJS..178...56F}. In addition, a full survey of OH maser
emission in the LMC has been performed \citep{2008MNRAS.385..948G}, as
well as photometric surveys in the optical
\citep[e.g.][]{2002ApJS..141...81M,2004AJ....128.1606Z}.

\subsection{Mid-infrared spectroscopic studies}

Several mid-infrared spectroscopic studies of representative targets
in object classes in the LMC have already been performed. Most studies
focus on individual objects or small samples, although a few
systematic studies of large samples exist. Here we provide an overview
of studies not including data from \emph{SAGE-Spec}.

\subsubsection{Silicates}
\label{sec:silicate}

Prior to the launch of \emph{Spitzer}, \citet{VWM_99_exgalxsil}
obtained 2--45 $\mu$m spectroscopy of R71, a Luminous Blue Variable in
the LMC, using the Short Wavelength Spectrometer
\citep[SWS;][]{GHB_96_SWS} on board the Infrared Space Observatory
\citep[\emph{ISO};][]{KSA_96_ISO}. This spectrum contained the first
detection of extragalactic crystalline silicates, and also showed the
presence of polycyclic aromatic hydrocarbons (PAHs).

With \emph{Spitzer}-IRS the possibility to obtain mid-infrared
spectroscopy of individual objects greatly expanded, allowing the
analysis of the dust mineralogy.  IRS spectroscopy of two B$[$e$]$
hypergiants (R126 and R66) revealed the presence of silicate dust,
where R66 also shows evidence for crystalline silicates and a dual
chemistry with the presence of PAHs \citep{2006ApJ...638L..29K}.  Both
of these sources are known to have disks, which may provide a suitable
environment for crystallization. Additional sources found with a
silicate mineralogy are IRAS 05003$-$6712, which shows the
characteristic features of crystalline silicates enstatite and
forsterite superposed on amorphous silicate features
\citep{ZMW_06_SpitzerLMC}, and HV 2310, a Mira-type star showing an
unusually-shaped 10 $\mu$m resonance, which suggests the presence of
both crystalline and amorphous silicates \citep{2006ApJ...638..472S}.

\subsubsection{Carbon-rich and oxygen-rich evolved stars}

The late-stage AGB stars that dominate the MSX 8-$\mu$m point source
list \citep{EVD_01_LMC} are almost exclusively carbon-rich, as is
expected in the low metallicity environment of the LMC
\citep[e.g.][]{ZMW_06_SpitzerLMC,2008ApJ...686.1056S}.  The C$_2$H$_2$
molecular absorption bands are deeper than those in their Galactic
analogs explained by a higher abundance of this species at low
metallicity \citep{ZMW_06_SpitzerLMC}. Particularly deep molecular
absorption bands are found in IRAS 04496$-$6958
\citep{2006ApJ...650..892S}, along with a possible detection of SiC at
11.3 $\mu$m in absorption.  Another object, SMP LMC 11, is classified
as a planetary nebula based on its emission lines although its
infrared spectral energy distribution is more reminiscent of a
post-AGB star \citep{2006ApJ...652L..29B}. The spectrum of SMP LMC 11
shows a variety of organic molecular bands in absorption, including
several first extragalactic detections. However,
\citet{2006MNRAS.371..415M} have performed a study of molecular bands
(C$_2$H$_2$ and HCN) in a larger sample of carbon stars and find that
the abundance of C$_2$H$_2$ is independent of metallicity, while the
HCN bands remain undetected.

An inventory of dust features in the IRS spectra of carbon-rich AGB
stars in the LMC is given by \citet{ZMW_06_SpitzerLMC}, and shows the
presence of SiC and MgS solid state components.
\citet{2008ApJ...681.1557L} have added a further 19 sources to this
sample, and devised a method to study the dust condensation sequence
in environments of differing metallicity.  It was also noted that the
occurrence of MgS correlates with diminished strength of the SiC
feature, suggesting that MgS forms a coating on SiC grains
\citep{LZS_07_carbonstars,2008ApJ...681.1557L}, although this seems
contradictory with observations of a sample of 7 of the most extremely
reddened carbon stars \citep{2008ApJ...688L...9G}, where the presence
of MgS apparently did not hamper the detection of SiC in absorption.

\subsubsection{Planetary Nebulae}

The prevalence of the carbon-rich phase in stellar evolution is
supported by evidence from IRS observations of Planetary Nebulae
(PNe). \citet{2007ApJ...671.1669S} have observed 41 PNe in both
Magellanic Clouds, 25 of which are LMC sources, and which were
previously observed using HST. Roughly half of the spectra of the LMC
PNe are dominated by atomic emission lines, while the remainder show
solid state features, of mostly carbon-rich species. Only two LMC PNe
show clear detections of oxygen-rich dust, in particular crystalline
silicates.  An independent study of atomic lines performed on a sample
of 25 LMC/SMC PNe yielded neon and sulfur abundances
\citep{2008ApJ...672..274B}, both of which are found to be lower than
the Galactic values, roughly in the same ratios as the metallicity
ratios with respect to the Milky Way.

\subsubsection{Supernovae and their remnants}

A detailed study of supernova remnant N132D by
\citet{2006ApJ...653..267T} included IRS observations and showed the
emission lines of $[$Ne{\sc iii}$]$ and $[$O{\sc iv}$]$, as well as
PAH emission features -- including relatively strong emission from the
PAH features at 14-20 $\mu$m.  The observations were of part of the
shell and a fast-moving knot. In supernova remnant N49 the PAH
features are less prominent compared to the atomic emission lines
\citep{2006AJ....132.1877W}, which may indicate PAH destruction by UV
radiation. The IRS spectrum of recent supernova SN 1987A is dominated
by emission from silicate dust, with a few atomic lines
\citep{2006ApJ...650..212B}. The dust mass derived is $\sim 2.6 \times
10^{-6} \, M_\odot$.

\subsubsection{Pre-Main Sequence stars}

IRAS 05328$-$6827 is the first YSO to be analyzed with the IRS in the
LMC, showing a CO$_2$ ice band that, compared to the H$_2$O ice band,
is deeper than that typically observed in the Milky Way
\citep{2005MNRAS.364L..71V}.  \citet{2009ApJ...699..150S} have
targeted YSO candidates suggested by \citet{2009ApJS..184..172G} and
spectrally identified 277 YSOs, thus greatly expanding the sample of
known YSOs in the LMC. These YSOs were subdivided in 6 different
groups based on the presence of CO$_2$ ice features, silicate
features, PAH features and atomic emission lines.

\subsubsection{Spectral catalogs of point sources}

A number of studies have compiled spectral catalogs. Spanning a range
of infrared colors, \citet{BKF_06_8micron} have selected and observed
60 of the brightest 8 $\mu$m sources, and classified 21 red
supergiants, 16 carbon-rich and 4 possible oxygen-rich AGB stars, 2
OH/IR stars and the 2 B$[$e$]$ stars discussed in
Sect.~\ref{sec:silicate}.  A smaller sample of 28 sources taken from a
range of pre-defined classes is presented by
\citet{2008ApJ...686.1056S} and reveals a veritable zoo of spectra.
The classification scheme devised by \citet{BKF_06_8micron} has been
used by \citet{2008AJ....136.1221K} to classify 250 of the most
luminous 8 $\mu$m MSX sources in the LMC, arriving at the conclusion
that in this flux-limited sample carbon-rich AGB stars indeed dominate
(35\% of the sources), closely followed by \ion{H}{2} regions (32\%;
some of which might contain massive YSOs), and at some distance Red
Supergiants (18\%). Less prominent in this sample are the populations
of oxygen-rich AGB stars (5\%), dusty early-type emission-line stars
(3\%) and foreground AGB stars (3\%) in this sample. The remaining 4\%
of sources could not be classified. This classification was tested by
\citep{2009AJ....138.1597B}, who found by studying the IRS spectra of
objects which were not previously classified in their earlier work
\citep{BKF_06_8micron}, that 22 out of 31 sources received the correct
classification.

\section{Observations}
\label{sec:observations}

\begin{deluxetable}{llc}
\tablenum{1}
\tablecaption{Summary of observations in the \emph{SAGE-Spec} program \label{tab:obsdata}}
\tablehead{\colhead{observing mode} & \colhead{number of targets} & \colhead{total obs. time}}
\startdata
IRS staring & 196 point sources & 108.7 hrs\\
IRS mapping & 10 atomic clouds & 64.2 hrs\\
            & 10 \ion{H}{2} regions &\\
            & 20 background &\\
MIPS SED & 10 atomic clouds & 20.5 hrs\\
                  & 10 \ion{H}{2} regions &\\
                  & 20 background &\\
MIPS SED & 48 point sources & 31.2 hrs\\
\enddata
\end{deluxetable}

The \emph{Spitzer} \emph{SAGE-Spec} program (PID: 40159) consists of
224.6 hours of spectroscopic observations of targets in the LMC
(Tab.~\ref{tab:obsdata}). The targets included point sources and
extended regions, both of which were observed using the IRS low
resolution and MIPS SED modes. Observations were done in the IRS
staring mode for 196 point sources, and 48 point sources were observed
in MIPS SED mode. In addition, 10 extended regions were mapped in both
the MIPS SED and IRS observing modes. These \emph{SAGE-Spec} data are
discussed in Sect.~\ref{sec:SAGEPS} (point sources) and
Sect.~\ref{sec:SAGEEXT} (extended regions).

In addition to the observations made as part of the \emph{SAGE-Spec}
program, we also deliver to the scientific community our new, homogeneous
reductions of all archival IRS and MIPS SED spectroscopic data within
the \emph{SAGE-LMC} footprint, as part of the \emph{SAGE-Spec}
legacy. Tab.~\ref{tab:archivalirs} lists the archival IRS staring mode
observations, while the archival IRS maps and MIPS SED observations
within the \emph{SAGE-LMC} footprint are discussed in
Sect.~\ref{sec:mipssedps} and Sect.~\ref{sec:mipssedext}.

\subsection{Point sources}
\label{sec:SAGEPS}

\subsubsection{Target selection}

\begin{figure} 
\plotone{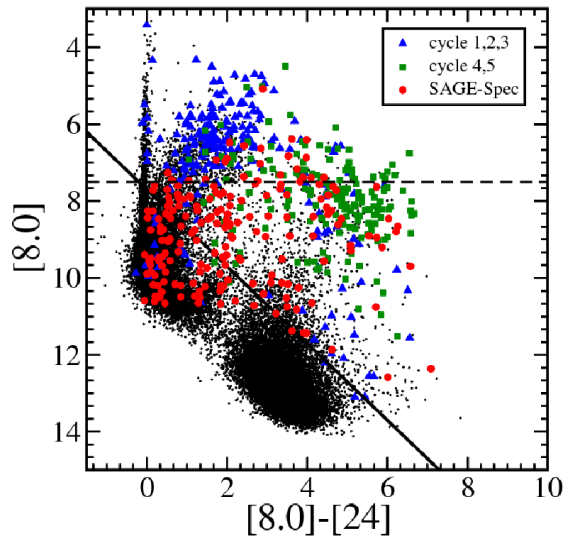}
\caption{[8.0] vs.~[8.0]-[24] color-magnitude diagram of infrared
point sources in the LMC. The black dots represent the sources for
which these colors are available in the \emph{SAGE-LMC} point source
catalog \citep{2006AJ....132.2268M}. Sources selected for observation
in the \emph{SAGE-Spec} program are indicated with red circles,
whereas archival targets observed in cycles 1--3, and cycles 4--5 are
shown as blue triangles and green squares, respectively. The latter
group is dominated by targets from PID 40650 \citep[PI:
Looney;][]{2009ApJ...699..150S}. The diagonal solid line in the
diagram represents the sensitivity limit that we applied to select
sources for LL observations. \emph{SAGE-Spec} sources below this line
were only observed with SL. The MSX detection limit at 8.0 $\mu$m is
shown as a horizontal dashed line.
  \label{fig:cmd}}
\end{figure}

Spectroscopic studies performed with \emph{Spitzer} prior to the
\emph{SAGE-LMC} survey, in observing cycles 1--3,
\citep[e.g.][]{BKF_06_8micron,ZMW_06_SpitzerLMC} have predominantly
targeted (extreme) AGB stars known before the launch of
\emph{Spitzer}. These objects are concentrated in the brightest part
of the [8.0] vs.~[8.0]-[24] color-magnitude diagram (blue triangles in
Fig.~\ref{fig:cmd}), above the MSX detection limit.

In order to explore the full life-cycle of dust in the LMC and to
classify completely the sources in the \emph{SAGE-LMC} photometric
catalogs, we have selected additional sources that cover the range in
luminosities and colors found in the \emph{SAGE-LMC} photometric
survey, focusing predominantly on the unexplored region below the MSX
detection limit, and including some additional bright, extremely red
sources (red circles in Fig.~\ref{fig:cmd}).  The \emph{SAGE-Spec}
program was executed in observing cycle 4, with the \emph{SAGE-LMC}
data becoming available to the \emph{SAGE-LMC}/\emph{SAGE-Spec} team
and the community prior to the proposal deadline for that
cycle. Several other proposals targeted point sources in the LMC below
the MSX detection limit, most notably the program \emph{An
Evolutionary Survey of Massive YSOs} \citep[PID: 40650; PI:
L.~Looney;][]{2009ApJ...699..150S}, in which about $\sim$300 candidate
YSOs were targeted, selected from the \emph{SAGE-LMC} observations
using independent photometry from \citet{2009ApJS..184..172G}; so this
area, too, received substantial coverage with IRS over the lifetime of
\emph{Spitzer}. The cycle 4 and 5 targets, dominated by the sample
proposed by \citet{2009ApJ...699..150S}, are marked with green squares
in Fig.~\ref{fig:cmd}.

For the \emph{SAGE-Spec} program, we arrived at a target list
containing 196 pointings, all of which have been observed using the
short low (SL) mode on IRS, while 128 of these were also observed in
the long low (LL) mode (see Tabs.~\ref{tab:irsss} and
\ref{tab:irsss2}). We focused on field stars, but also included a
small sample of objects from clusters with known metallicities and
ages, yielding targets of a much better constrained pedigree than
field stars. Cluster stars were mostly overlooked prior to the
\emph{SAGE-Spec} survey. The cluster sources are indicated as such in
Tab.~\ref{tab:irsss}.

\begin{figure} 
\plotone{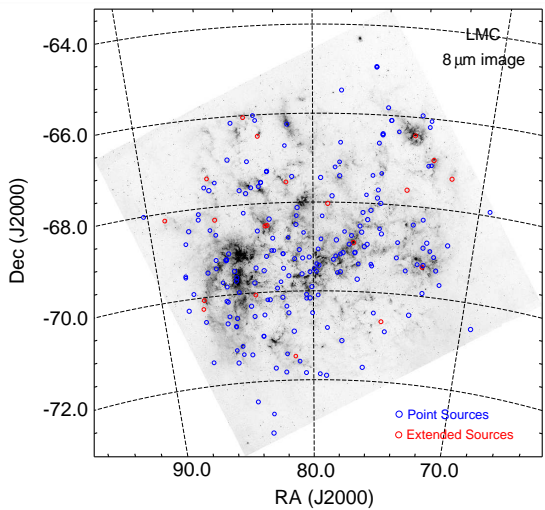}
\caption{Distribution of \emph{SAGE-Spec} IRS targets over the
LMC. The blue symbols indicate the positions of point sources targeted
with the IRS staring mode, while the red symbols indicate the central
positions of the 20 MIPS SED/IRS maps of extended regions. The
positions are overplotted on the IRAC 8-$\mu$m footprint of the
\emph{SAGE-LMC} survey \citep{2006AJ....132.2268M}
\label{fig:distr}}
\end{figure}

We selected candidates in a range of object classes. The sample
includes candidate AGB stars, both O-rich and C-rich, selected from
the work by \citet{2009AJ....137.4810S}; and potential YSO sources
taken from the list of candidates selected on their colors and
magnitudes \citep{2008AJ....136...18W}. Due to the overlapping
color-magnitude space with YSOs, we also expected to detect background
galaxies \citep{BMO_06_evolved}. In addition, we ensured
representation of rarer objects, such as post-AGB stars and PNe using
additional criteria. For the post-AGB stars the samples of
\citet{1998AJ....115.1921A} and \citet{WC_01_LMC} provided a starting
point; while the PNe were drawn from lists of LMC PNe assembled by
\citet{1997A&AS..121..407L} and \citet{2006MNRAS.373..521R},
representing an adequate sampling of electron density and temperature,
morphology, and infrared colors.  In order to fully cover the
color-magnitude space parametrized by IRAC, MIPS and 2MASS magnitudes,
we selected a total of 13 sources from under-represented regions, such
as the region defined by $[3.6]-[8.0] > 5$ and the region bordered by
$[8.0] > 9$ and $J > 13.3$.  The distribution of the selected sources
over the LMC is shown in Fig.~\ref{fig:distr}.

\paragraph{MIPS SED point source target selection}

We observed 48 point sources with MIPS SED (Tables~\ref{tab:irsss} and
\ref{tab:archivalirs}), which is only feasible for the brightest
objects in the LMC. A minimum flux level of 100 mJy at 70 $\mu$m is
required to obtain a S/N of $\sim$3 for $20\times10$s integrations. We
also required all targets selected for MIPS SED observations to have
been observed with IRS, either within the context of the
\emph{SAGE-Spec} program, or archival programs in cycle 1--3.

\subsubsection{IRS staring mode}
\label{sec:IRSstaring}

\paragraph{Observations}

The IRS observations of the point sources in the sample are carried
out in staring mode. All 196 selected targets were observed using
IRS-SL, while the 128 targets with a MIPS-24 $\mu$m flux $\gtrsim 5.7$
mJy were also observed with IRS-LL. Each target was observed in both
nod positions, located one-third and two-thirds along the
slit. Integration times were chosen based on the IRAC-8.0 and MIPS-24
$\mu$m flux levels, and targeted to result in a S/N of 60 in SL and 30
in LL, in principle sufficient to analyze and classify dust features
on top of a stellar continuum.

Although we observed toward 196 positions, one observation (AOR key
22402560) clearly showed the contribution of two different objects in
SL and LL; at short wavelengths the spectrum is due to GV 60, present
at the observed location, while the LL spectrum is dominated by nearby
Wolf-Rayet star LH$\alpha$ 120--N 82.

\paragraph{Data reduction SAGE-Spec data} The processing of data for
the \emph{SAGE-Spec} program began with the flat-fielded images
produced by the S18.7 version of the data-reduction pipeline at the
\emph{Spitzer} Science Center (SSC).  To extract spectra from the
flat-fielded images and calibrate them spectrophotometrically, we
followed the procedure used by several other programs in the
Magellanic Clouds and elsewhere in the Local Group
\citep[e.g.~][]{2006ApJ...645.1118S,ZMW_06_SpitzerLMC,LZS_07_carbonstars,2007MNRAS.382.1889M,2008ApJ...686.1056S,2009Sci...323..353S,2009MNRAS.396..598L}. The
SL and LL modules each have two apertures, one for the second-order
data covering the shorter-wavelength portion, and the other for the
first-order data covering the longer-wavelength portion.  When the
target is in the second-order aperture (SL2 or LL2), a short piece of
first-order data is also obtained, which is referred to as the bonus
order (SL-bonus or LL-bonus).  The bonus order provides overlap with
the true first-order data (SL1 or LL1), making it possible to correct
the spectra for discontinuities between the orders.

Observations were constructed so that the number and length of
integrations in each aperture within a module matched, giving us
flexibility on the background subtraction method.  Generally, in SL,
we chose as the background for a given exposure the corresponding
exposure with the target in the other aperture.  These aperture
differences place the positive and negative beams about 79$''$ apart,
compared to 19$''$ if we had used nod differences.  In SL, nod
differences would have placed the positive and negative beams close
enough to each other to interfere for extended or complex sources.
Consequently, we only reverted to nod differences for those
observations where the background emission showed a gradient over the
79$''$ throw.  For LL, the default for background subtraction was a
nod difference, which placed the positive and negative beams
$\sim$56$''$ apart.  We generally avoided using aperture differences
in LL because the beams were 192$''$ apart, which would often expose
us to the more severe background gradients present in the LMC longward
of 15~$\mu$m.  We examined each image and spectrum carefully to assess
when it was necessary to deviate from the default
background-subtraction method to avoid either additional sources or
complex backgrounds.  In some cases, we even reverted to using the
image with the target in the other aperture and nod as the background
(a cross difference).

In addition to removing the background, differencing the data also
corrects most of the \emph{rogue} pixels in an image.  These pixels
exhibit dark currents different than their usual levels, but generally
stable for the duration of a given observation.  Some pixels
remain problematic for a variety of reasons.  Most are flagged as such in the
rogue pixel masks provided by the SSC.  We built \emph{super-rogue}
masks assuming that up to the campaign in which a target was observed,
a pixel could be defined as bad if it had been flagged as bad in two
previous campaigns.  We replaced all flagged pixels using the {\tt
imclean} algorithm developed at Cornell and distributed as a part of
{\tt irsclean} by the SSC\footnote{\tt
http://ssc.spitzer.caltech.edu/dataanalysistools/
tools/irsclean/}.
This algorithm replaces bad pixels by comparing the point-spread
functions (PSFs) in adjacent rows.

To extract spectra from the differenced and cleaned images, we used
the SSC pipeline modules {\tt profile} (to locate the source in the
slit), {\tt ridge} (to map the source position and extraction aperture
in the image), and {\tt extract}.\footnote{These modules are available
in {\rm SPICE}, the \emph{Spitzer} IRS Custom Extraction package.}  The
{\rm extract} module extracts a spectrum from an image by summing the
flux within a pseudo-rectangle defined for each wavelength element.
When the boundaries of a pseudo-rectangle cross a pixel, the flux is
assumed to be evenly distributed within that pixel.  The
pseudo-rectangles are centered on the center of the PSF at each
wavelength element, and their width increases proportionally with
wavelength.  The tapered-column extraction within SMART
\citep{HDH_04_smart} was designed to follow this algorithm precisely,
and it gives very similar results .

\emph{Spitzer} obtains IRS data in a series of Data Collection Events
(DCEs).  We extracted spectra separately from each DCE, then co-added
them to produce one spectrum per nod position.  This step produces a
mean flux density and a standard deviation, which we divided by the
square root of the number of DCEs to estimate the uncertainty in flux
density.  To calibrate the co-added spectrum from each nod position,
we determined spectral corrections using IRS observations of the
standard stars HR~6348 (K0 III), HD~166780 (K4 III) and HD~173511 (K5
III).  HR~6348 served as the standard for SL (to avoid any
difficulties with the strong SiO absorption in the later K giants),
while all three served as standards for LL (to maximize the S/N).  We
chose to use K giants rather than $\alpha$~Lac \citep[A1\,V;
e.g.][]{2006ApJS..165..568F} because of the difficulty in predicting
the strength of the hydrogen recombination lines in the low-resolution
modules.

When combining the spectra from the two nod positions for a given
order, we replaced the uncertainty when a comparison of the two
spectra produced a larger value.  When the uncertainty for a given
pixel exceeded the average uncertainty in the neighborhood by a factor
of five (typically), we used only the data from the nod which were
closer to neighboring data.  This spike-rejection algorithm removed
the occasional spikes and divots which survived the cleaning step
above.

\begin{deluxetable}{lc}
\tablenum{2}
\tablecaption{Wavelength ranges \label{tab:orders}}
\tablehead{\colhead{segment} & \colhead{wavelength ($\mu$m)}}
\startdata
SL2      &  5.10--7.59\\
SL-bonus &  7.23--8.39\\
SL1      &  7.59--14.20\\
LL2      & 13.95--20.54\\
LL-bonus & 19.28--21.23\\
LL1      & 20.46--37.00\\
\enddata
\end{deluxetable}

Finally, spectra from the six orders were combined into one spectrum
(SL2, SL-bonus, SL1, LL2, LL-bonus, LL1; see Table~\ref{tab:orders}).
First the bonus-order data were averaged with the first- and
second-order data where they overlapped and were within the defined
range of valid data.  Then the spectra were stitched together to
remove discontinuities between segments.  These discontinuities arose
primarily from mispointings, almost always in SL, with its narrower
slit (3.6$''$ vs.~10.0$''$).  In general, we assumed that the
corrections were always upward to the best-centered spectral segment.
All corrections were multiplicative and scalar (i.e.\ not a function
of wavelength).  To conclude the processing, we trimmed the spectra of
those portions at the ends of each segment which proved impossible to
calibrate reliably.  We also reset uncertainties which indicated a
signal/noise ratio $>$ 500, as these values are unlikely and can
adversely affect algorithms which use the S/N to weight the data.

\paragraph{Archival data}

We have perused the archive for all staring mode observations within
the \emph{SAGE-LMC} footprint, and complemented this list with a few
mini-map observations, apparently designed to target point
sources. The observations are listed in
Tab.~\ref{tab:archivalirs}. Most observations are single pointing
staring mode observations, but in some cases more complex settings are
used. Examples are the mini-maps, which appear to be designed to cover
the entire point-spread-function, and cluster mode observations, where
several pointings are strung together in a single AOR. All archival
targets were reduced following the scheme described above.  One target
was only observed with SL2, forcing us to depart from our default
background subtraction using aperture differences, since there is no
other aperture to extract. For those data, and in other cases where
complex backgrounds made aperture differences inadvisable, we used nod
differences in SL.

\begin{figure} 
\plotone{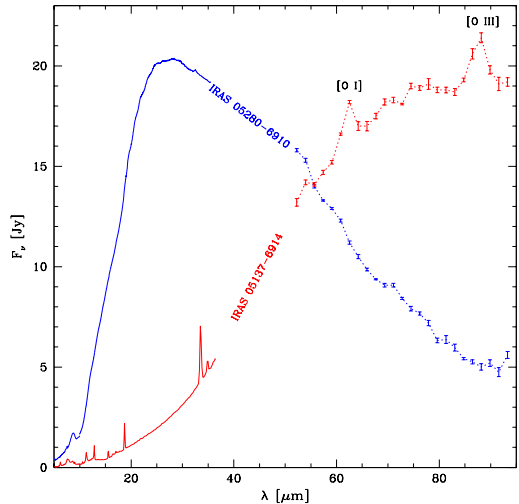}
\caption{Combined MIPS SED \citep{2010AJ....139...68V} and IRS view of
two bright, compact IR sources: one, OH/IR star IRAS\,05280$-$6910 is
dominated by a warm circumstellar dust envelope with the 10 and 18
$\mu$m silicate features in absorption and a declining, featureless
continuum at far-IR wavelengths; the other, ultra-compact \ion{H}{2}
region IRAS\,05137$-$6914 is dominated by cold dust and an
emission-line spectrum both at mid- and far-IR wavelengths.
\label{fig:specps}}
\end{figure}

We include both low-resolution (SL and LL) and high-resolution (SH and
LH) in the final data delivery.  Both Short-High (SH) and Long-High
(LH) have short slits, which limit the background-subtraction method.
Later in the \emph{Spitzer} mission, the SSC strongly recommended that
all SH and LH observations include dedicated background observations,
but most observations early in the mission came with no background
observations.  Where these were available, we subtracted them from the
images before extracting.  When they were not, then we were forced to
skip this important step and continue without background subtraction.
In all cases, we performed a full-slit extraction, summing all of the
flux in the slit at each wavelength.  When stitching high-resolution
data, we have not applied different corrections to the orders within
SH or LH, because these were all obtained simultaneously and cannot
differ due to pointing effects.  Sample spectra showing the reduced
IRS staring mode data are shown in Fig.~\ref{fig:specps}.

\subsubsection{MIPS SED point sources}
\label{sec:mipssedps}

\paragraph{Observations}

For the MIPS SED mode observations, the integration times are
determined based on the measured 70~\micron\ emission, from
\emph{SAGE-LMC}, and the slope of the SED.  The chop size was chosen
to place the background measurement in a region relatively free of
emission for the range of dates expected for the observations.  Due to
the high efficiency of \emph{Spitzer} scheduling, the MIPS SED
observations were taken earlier than expected, resulting in some chop
regions not being taken in the ideal locations.  The chop sizes were
normally $1^\prime$, except in five cases were a chop of either
$2^\prime$ or $3^\prime$ was employed. Integration times ranged
between 24 s and 200 s.

For the LMC, there is one additional point source with MIPS SED
observations available in the \emph{Spitzer} archive.  This is
SN~1987A and has been observed as part of programs 30067, 40149 \&
50444 (PI: Dwek).  The IRS spectroscopy and MIPS and IRAC photometry
associated with these programs is already published
\citep{2006ApJ...650..212B}.

\paragraph{Data Reduction}

The MIPS SED point source observations were reduced using the MIPS DAT
\citep[see~][]{2005PASP..117..503G}.  The extraction was done for a 5
pixel wide aperture centered on the collapsed profile maximum.  For
the majority of the sources, the off-source chop observations were
used to do the initial background subtraction (except when there was
significant emission in the off-source chop position).  In addition,
background subtraction was done by subtracting the measurements made
with the same extraction aperture in a region near the edge of the
slit.  A wavelength dependent aperture correction was applied to
extrapolate to infinite extraction aperture.  Finally, a smoothed
sensitivity function was applied to convert from instrumental to
physical units.  The uncertainty due to the absolute calibration is
15\% \citep{2008PASP..120..328L}. Fig.~\ref{fig:specps} shows the MIPS
SED spectrum of point sources IRAS 05280$-$6910 and IRAS 05137$-$6914,
as examples of the quality of the data. More details on, and first
science results from, the MIPS SED observations of point sources in the
LMC can be found in \citet{2010AJ....139...68V}.

\subsection{Extended regions}
\label{sec:SAGEEXT}

\subsubsection{Target selection}

\begin{figure} \plotone{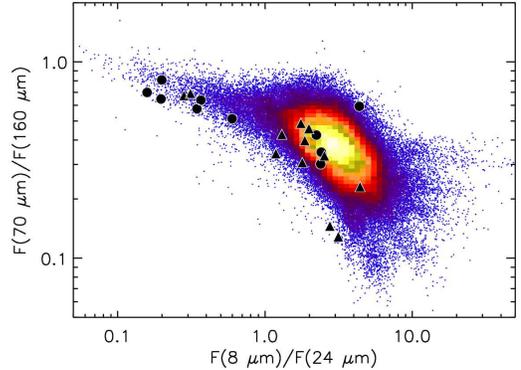}
\caption{IRAC/MIPS color-color diagram for extended regions. The
\emph{SAGE-LMC} data for all three MIPS bands and the IRAC 8-$\mu$m
band were convolved to the spatial resolution of the MIPS-160 $\mu$m
band. The background dots represent the ratios of these bands for each
MIPS-160 spatial resolution element. The \ion{H}{2} regions that were
observed for the \emph{SAGE-Spec} program are plotted as circles and
the diffuse regions as triangles.
\label{fig:ext}}
\end{figure}

For the interstellar extended source observations, we distinguish
between highly irradiated (ionized) extended regions (i.e. \ion{H}{2}
regions) and lower-level irradiated clouds, described as atomic and
molecular clouds. The \ion{H}{2} regions in the \emph{SAGE-Spec}
program (see Tab.~\ref{tab:ext}) have been selected on size and the
r.m.s.~density as derived from the H$\alpha$ emission
\citep{1986ApJ...306..130K}.  The source list covers a range in sizes
from $1'$ to $16'$, while the electron density range spans about two
orders of magnitude. H\,{\sc ii} regions cover only part of IRAC/MIPS
color-color space in the LMC (circles in Fig.~\ref{fig:ext}).  The
atomic regions were selected from the $F_{8.0}/F_{24}$ and
$F_{70}/F_{160}$ IRAC/MIPS color-color space, as observed by
\emph{SAGE-LMC}, and are chosen to span a wide range of these IRAC and
MIPS colors (Fig.~\ref{fig:ext}). Based on the overall pixel
statistics, we divided the color-color space into a number of
canonical bins (e.g.~$5< F_{8.0}/F_{24} < 10$ and $F_{70}/F_{160}
<0.1$), and selected regions with high concentrations of pixel values
in these bins. As an additional criterion we required these regions to
have a distinct identity, for instance a feature in the Spitzer maps
or a CO or \ion{H}{1} cloud. The selection is somewhat exploratory in
nature due to its small size.

While the \ion{H}{2} regions in general tend toward $F_8/F_{24}$ ratio
$\sim$ 1, the atomic regions chosen span higher ratios ranging from 1
to 10, on average. This selection does not guarantee that \ion{H}{2}
regions are not present in the atomic regions mapped, but samples
regions which would be dominated by properties characteristic of
atomic ISM.  The \emph{SAGE-Spec} sample has been extended by the
archival IRS and MIPS SED maps of the 30 Doradus \ion{H}{2} region
\citep{2009ApJ...694...84I} and two additional atomic regions in the
LMC from PID 40031 (PI: G.~Fazio; SSDR 11 and SSDR 12 in
Tab.~\ref{tab:ext}), of which the data remain unpublished so far.  The
central positions and sizes of all 23 extended regions are listed in
Tab.~\ref{tab:ext}.

\subsubsection{IRS mapping mode}
\label{sec:irsmapping}

\paragraph{Observations} For the atomic and molecular cloud
observations, our sensitivity objective was to obtain spectral maps
such that when spatially integrated over a $1\arcmin \times 1\arcmin$
region, we would achieve a S/N = 10.  We used exposure times per pixel
of $4 \times 14$s (SL) and $4 \times 30$s (LL) and spectral mapping to
cover each $1\arcmin \times 1\arcmin$ region. All selected \ion{H}{2}
regions are mapped in strips that have a width of $1\arcmin$, and the
length being the diameter of the \ion{H}{2} region.  The mapping is
done in such a way that the SL slit is stepped in the cross-slit
direction by the diameter of the \ion{H}{2} region and contains 2
pointings in the slit direction. The LL slit is stepped in the
cross-slit direction by $1\arcmin$ and again in the slit direction by
the diameter of the \ion{H}{2} region, thus obtaining a $1\arcmin$
wide strip in both LL and SL. There are four 6s exposures for both
modules, and, since the maximum length of an IRS AOR is 6 hours, the
total length of the strip is limited to 5.4$'$. The largest \ion{H}{2}
regions in our sample are therefore not mapped to their full diameter.
For both the low surface brightness clouds and the \ion{H}{2} regions,
dedicated off-source observations were obtained to remove the time
dependent IRS detector hot pixels and zodiacal light background.

\paragraph{Data Reduction}

We used the standard pipeline data as produced by the SSC.  The
individual observations were combined into a spectral cube using
CUBISM \citep{2007PASP..119.1133S}, and these spectral cubes were
merged together using custom software \citep{2009ApJ...696.2138S}.
Each independent spectrum in the cube was fit using PAHFIT
\citep{2007ApJ...656..770S} after convolution to a common resolution
using custom convolution kernels
\citep{2008ApJ...682..336G,2009ApJ...696.2138S}. The fit parameters
are used to construct the feature maps.  For the molecular and atomic
regions, we did achieve a S/N of 10, especially at $\lambda > 10$
$\mu$m. For the \ion{H}{2} regions, more mixed results emerged, with
only 6 of the 10 regions meeting the S/N goal.  In those regions with
a S/N $>10$ it is possible for us to investigate spatial variations in
the properties of dust and PAHs, and correlate this with the
interstellar radiation field measured through PAH feature strengths
and atomic line ratios.

\begin{figure} 
\plotone{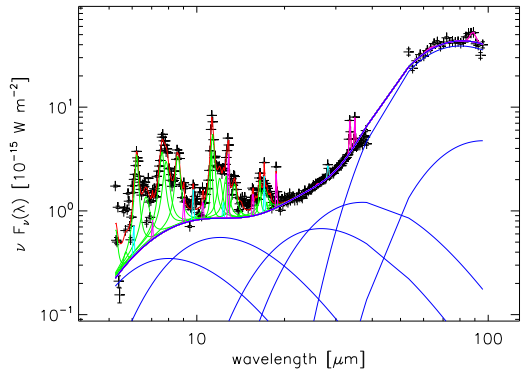}
\caption{Integrated IRS and MIPS SED spectrum of SSDR 1. The spectrum
is spatially integrated over a circle with diameter 60$"$, centered on
the position listed in Tab.~\ref{tab:ext}. A modified form of the
PAHFIT model has been applied (see main text). The pluses indicate the
data, and the solid lines are the various components of the PAH fit
model, with blue representing the dust continuum, cyan the atomic
transitions, and green the PAH bands. The overall fit is indicated
with a red line.
  \label{fig:atomicspec}}
\end{figure}

We have also spatially integrated the IRS (and MIPS SED) spectroscopy,
as Fig.~\ref{fig:atomicspec} shows for one of the atomic regions. The
integrated SED of this region was extracted from the individual IRS
order cubes and MIPS SED cube over the region in common between all
the observations.  In this example, the extracted region was a 60$''$
diameter circle.  When studying spatial variations is not possible due
to low S/N, the integrated spectra still yield global information on
the dust and PAH properties, and the radiation fields in these
environments.

\subsubsection{MIPS SED}
\label{sec:mipssedext}

\paragraph{Observations}

For the \ion{H}{2} regions, the MIPS SED observations roughly coincide
with the peak of the SED and pick up any strong [\ion{O}{1}]
63~\micron\ and [\ion{O}{3}] 88~\micron\ lines
\citep{2009ApJ...694...84I}. For the atomic and molecular regions the
MIPS SED observations constrain dust temperatures, and, in particular,
the very small grain emission properties. All \emph{SAGE-Spec}
extended regions are mapped with 1/2 slit offsets in both slit
dimensions (9$''$ cross-slit and 1.25$''$ along-slit), with the
minimum exposure time of 3~s. With these exposure times, we aimed to
achieve a S/N of 5 per spatial bin for the \ion{H}{2} regions, while for the
diffuse regions, the objective is for a S/N of 5 for the spatially
integrated SED ($1\arcmin \times 1\arcmin$ region).  Indeed, these S/N
goals were met for 7 out of 10 \ion{H}{2} regions, and all of the
diffuse regions. For all extended regions, dedicated background
observations off the LMC were obtained.

For the LMC, there are two additional extended sources with MIPS SED
observations available in the \emph{Spitzer} archive.  They are 30~Dor
\citep{2009ApJ...694...84I}; and the 70~\micron\ excess \citep[as
described by][]{2008AJ....136..919B} region observed in PID 40031, for
both of which IRS mapping mode observations are also available (see
Sect.~\ref{sec:irsmapping}).

\paragraph{Data Reduction}

The MIPS SED extended source observations were reduced using the MIPS
DAT v3.10 \citep{2005PASP..117..503G}, in a way similar to MIPS
imaging data \citep[see, e.g.,][]{2007ApJ...655..863D} and calibrated
according to the prescription of \citet{2008PASP..120..328L}.  Using
the MIPS DAT we constructed three different products: on-source
background subtracted, on-source only, and off-source only rectified
mosaics combining all the appropriate observations in an AOR.  The
dedicated off-LMC background observations were subtracted from the on-
and off-source mosaics

For each spectral map, and using the on-source and off-source MIPS DAT
products, spectral cubes were populated by assuming that the slit is 2
pixels wide and has the coordinates and orientation found in the
header of each input image.  For any pixel in the output cube where
multiple input values are available, the output value is the mean of
the input values, weighted by the inverse square of the uncertainty
associated with that value.  This procedure has been captured in
custom IDL software, which parallels that of the similar software for
IRS spectra \citep[CUBISM;][]{2007PASP..119.1133S}.  An example of the
resulting spectra, integrated over a larger region, is shown in
Fig.~\ref{fig:atomicspec}.

\subsection{Data products and dissemination}
\label{sec:delivery}

As part of the \emph{SAGE-Spec} Legacy program we deliver basic and
enhanced data products to the astronomical community\footnote{The
\emph{SAGE-Spec} data products are available on {\tt
http://data.spitzer.caltech.edu/popular/sage-spec/}. The unprocessed
data are also available through the SSC archive tool Leopard}.
Reduced spectra for all point sources and extended sources within the
\emph{SAGE-LMC} footprint are included in the delivery in the form of
tables and plots. This dataset includes both the \emph{SAGE-Spec} and
archival observations within this footprint. The enhanced data
products include a spectral catalog of the archival and the SAGE-Spec
point sources, along with a source classification. In addition, a
photometric classification scheme will be derived, and applied to the
entire \emph{SAGE-LMC} point source catalog.

\begin{figure} 
\plotone{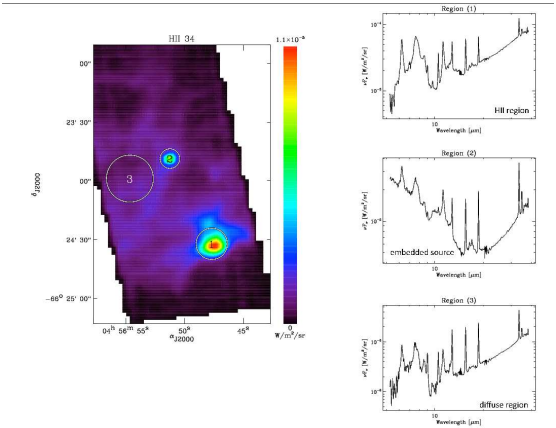}
\caption{Example of a spectral map of an H~{\sc ii} region (LHA 120-N
11; DEM L 34) and its surroundings. The map shows the integrated
intensity between 6 to 9~$\mu$m for the region where there is full
coverage over the entire IRS LL and SL range. The panels on the right
show the extracted spectra from different regions in the map. Region 1
is centered on N11B, a compact \ion{H}{2} region. Its IRS spectrum
shows nebular line and PAH emission on a steeply rising IR
continuum. Region 2 is centered on an embedded cluster which shows a
strong stellar continuum together with an IR excess due to the
surrounding dust. Region 3 represents a more diffuse region and it
exhibits very prominent nebular lines in its IRS spectrum. (Hony et
al.~\emph{in prep.})
\label{fig:hiispec}}
\end{figure}

In the cases where spectral maps were performed on extended sources,
the data will be spatially averaged into one spectrum.  We also
deliver spectral data cubes for spectral maps.  From those data cubes,
maps in selected features or spectral lines will be provided for
extended regions with sufficient flux in the intended
features. Fig.~\ref{fig:hiispec} shows an example of these spectral
feature maps.

  The first delivery took place on 1 August, 2009, and included the
reduced data of all \emph{SAGE-Spec} IRS staring mode observations,
along with a first batch of MIPS SED point source observations
\citep{SAGESpec-delivery2}.  The second delivery in March 2010
contains the reduced spectral data cubes for the extended regions
observed with MIPS SED and IRS (both \emph{SAGE-Spec} and archival
data) and the reduced archival IRS staring mode spectroscopy from
Cycles 1--3 \citep{SAGESpec-delivery2}.  Two more deliveries are
planned at $\sim$6 month intervals, encompassing archival data from
Cycles 4 and 5, and the enhanced data products.

\section{First results}
\label{sec:results}

Here, we list some first scientific results of the \emph{SAGE-Spec}
collaboration, in the context of both program aims: Following the life
cycle of gas and dust (Sect.~\ref{sec:lifecycle}) and the
classification of point sources (Sect.~\ref{sec:pointsources}).

\subsection{Life cycle of gas and dust}
\label{sec:lifecycle}

\subsubsection{Evolved stars}

\paragraph{Carbon-rich post-AGB stars}

\begin{figure}
\plotone{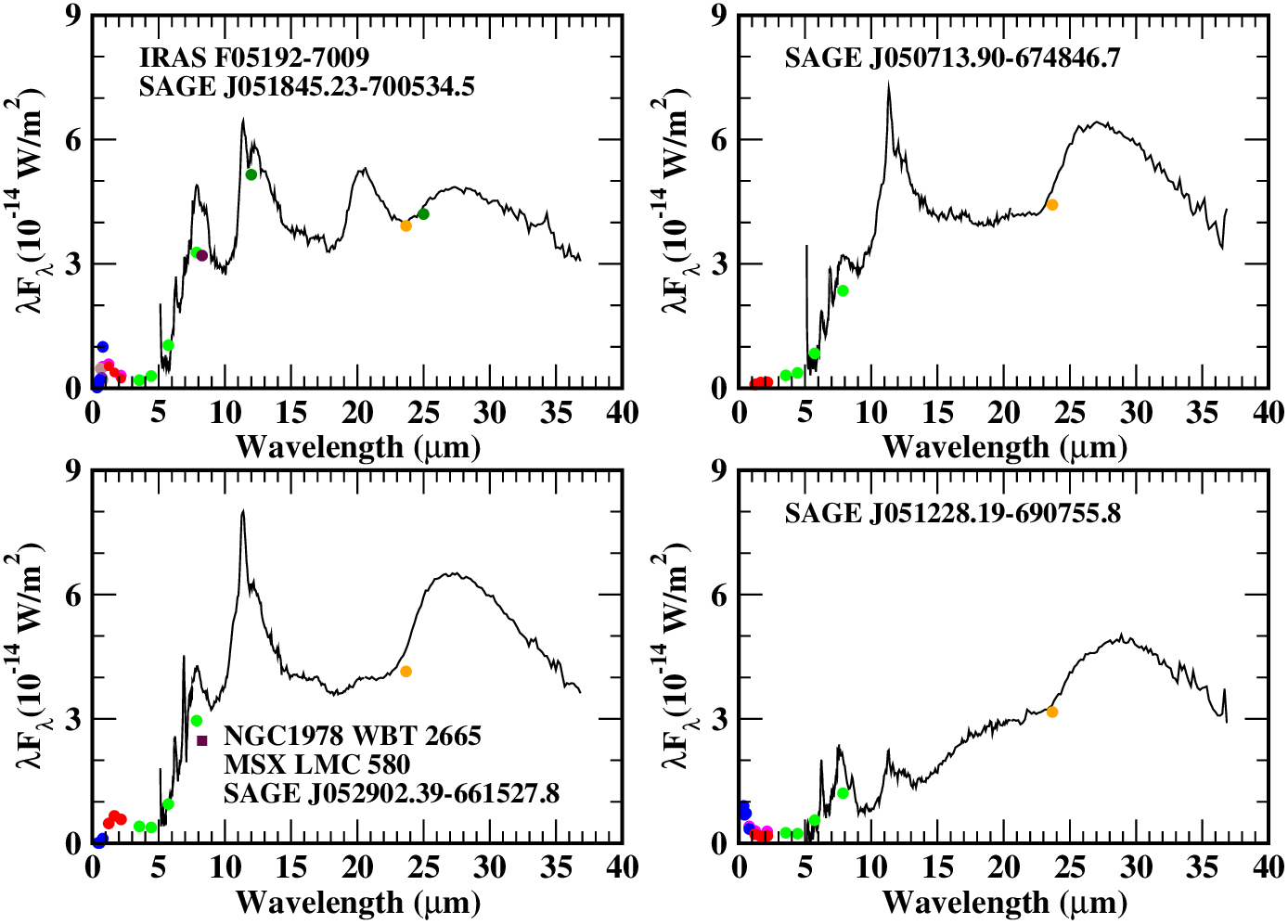}
\caption{IRS spectra of the four post-AGB candidate objects are given
all on the same scale for direct comparison.  In each panel the
available photometry from the literature are also plotted: green for
the \emph{SAGE-LMC} IRAC photometry, orange for \emph{SAGE-LMC}
MIPS-24 photometry, red for 2MASS photometry
\citep{2006AJ....131.1163S,2003AJ....126.1090C}, maroon for MSX
\citep[from the reject catalog for NGC 1978 WBT
2665;][]{2004AJ....128..889P}, blue for the Magellanic Cloud
Photometric Survey photometry \citep{1997AJ....114.1002Z}, magenta for
DENIS photometry \citep{2000A&AS..141..313F}, dark green for IRAS
photometry \citep{BNH_88_IRASPSC}, and brown/cyan for Palomar Sky
Survey photometry \citep{2003AJ....125..984M}.  No scaling has been
applied to any of the data values.
\label{fig:ppn}}
\end{figure}

Four carbon-rich post-AGB objects in the combined \emph{SAGE-Spec} and
archival samples can easily be identified by their spectral
characteristics: they have strong PAH emission and exhibit the 30
$\mu$m feature generally attributed to MgS grains. In addition, the
dust temperature is clearly low compared to that in carbon-stars still
on the AGB as judged from the shape of the continuum.  These objects
most likely have left the AGB within the last few hundred years and
are evolving to become planetary nebulae.  The IRS spectra of the four
objects along with the available photometry in the literature are
presented in Figure~\ref{fig:ppn}.  Three of the objects are
\emph{SAGE-Spec} targets and selected as likely post-AGB candidates,
while object NGC 1978 WBT 2665 was part of the sample of PID 3591
(Tab.~\ref{tab:archivalirs}), and selected on its infrared colors
using the classification by \citet{EVD_01_LMC}.

IRAS F05192$-$7009 shows a strong 21 $\mu$m feature, so far only
observed in post-AGB objects in our Galaxy \citep[see for example][and
references therein]{2009ApJ...694.1147H}, thus representing the first
extragalactic detection of this feature.  In the other three objects
it is weak or absent. For object SAGE1C J051228.18$-$690755.7 the data
around the position of the feature possibly contain an artefact, so it
is possible that there is a 21 $\mu$m feature present, but if so it is
quite weak and it seems more likely that the feature is absent.
Preliminary dust radiative transfer modeling for IRAS 05192$-$7009
shows that that the 21 $\mu$m and 30 $\mu$m feature shapes are the
same as those observed in Galactic objects, and suggests that the bulk
of the dust in the circumstellar shell is hotter than that found for
the Galactic 21 $\mu$m sources.  The models indicate that the dust
shell for IRAS 05192$-$7009 is somewhat more massive than those of a
number of the well studied Galactic 21 $\mu$m objects such as IRAS
07134$+$1005 and IRAS 22272$+$5435.  The same may be the case for all
four objects which would then suggest that they have left the AGB more
recently than most of or all of the Galactic 21 $\mu$m objects.

Compared to Galactic objects of this type, the LMC sources show very
strong PAH emission features including the rarely seen 6.9 $\mu$m
feature.  Comparison of the IRAC colors of these objects to the
simulated IRAC/MIPS colors of Galactic 21 $\mu$m objects with good
quality ISO SWS data indicates that the [3.6] $-$ [8.0] colors for the
LMC objects are larger than those for the Galactic objects, which may
be due to some combination of being less evolved off the AGB and
having the strong PAH features. All of these objects have a well
defined position in a K $-$ [8.0] vs.~K $-$ [24] color-color diagram
which suggests that further candidate post-AGB objects can be
identified from the \emph{SAGE-LMC} data.

\paragraph{RV Tauri stars}

A specific sub-class of post-AGB stars is formed by the RV\,Tauri
stars. These are pulsating stars which occupy the high-luminosity tail
of the population II Cepheids. Their light curves are characterized by
a succession of deep and shallow minima. Many objects also show
significant cycle-to-cycle variability.

One of the more remarkable properties of RV\,Tauri stars is that the
observed chemical pattern in the photospheres of many Galactic
RV\,Tauri stars is the result of a chemical rather than a
nucleosynthetic process: The photospheres are found to be deficient in
refractory elements (like Fe and Ca and the s-process elements), while
the non-refractory elements are not (or much less) affected
\citep[][and references therein]{giridhar05, maas05}. The photospheric
patterns can be understood by a process in which gas-dust separation
is followed by re-accretion of only the gas, which is poor in
refractory elements. This process has likely only taken place in
systems which are surrounded by stable dusty disks
\citep{1992A&A...262L..37W}, which RV\,Tauri stars are known to
possess \citep{V_03_postAGB}.

These dusty disks around evolved objects are ideal environments to
foster strong grain processing and in a \emph{Spitzer} survey of 21
Galactic sources \citet{2008A&A...490..725G} showed that very high
crystallinity prevails, and is dominated by magnesium-rich end members of
olivine and pyroxene silicates.  RV\,Tauri stars in the LMC were found
by the microlensing survey MACHO \citep{1998AJ....115.1921A} and high
resolution optical spectroscopy revealed that depletion of refractory
elements is also observed in their photospheres
\citep{2007A&A...463L...1R}.

\begin{figure} 
\plotone{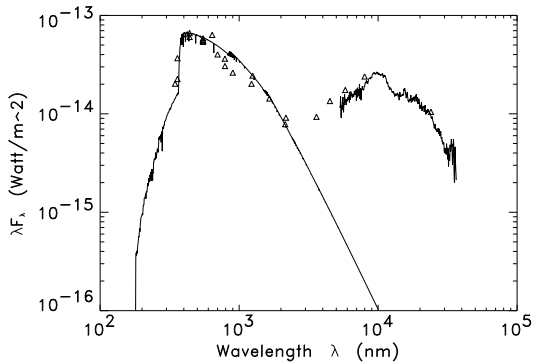}
\caption{The de-reddened SED of RV\,Tauri star MACHO 82.8405.15
\citep{2009A&A...508.1391G}. Triangles are broad-band fluxes. The full
line is the associated Kurucz stellar atmosphere model. The
\emph{SAGE-Spec} infrared spectrum is also shown. Thermal emission
from amorphous silicates dominates the spectrum but substructure in
the 10 $\mu$m profile as well as the presence of smaller features at
longer wavelengths indicate the presence of crystalline silicates
\citep{2009A&A...508.1391G}. This object shows a clear correlation
between abundance pattern \citep{2007A&A...463L...1R} and condensation
temperature of species, consistent with re-accretion.
  \label{fig:rvtau}}
\end{figure}

Using \emph{SAGE-Spec} data, we showed that also in the LMC, the
RV\,Tauri stars have stable disks, rather than dusty outflows
\citep{2009A&A...508.1391G} and the close connection between
photospheric depletion and the stable dusty environment is illustrated
in Fig.~\ref{fig:rvtau}.

\subsubsection{Interstellar medium}

\begin{figure} 
\plotone{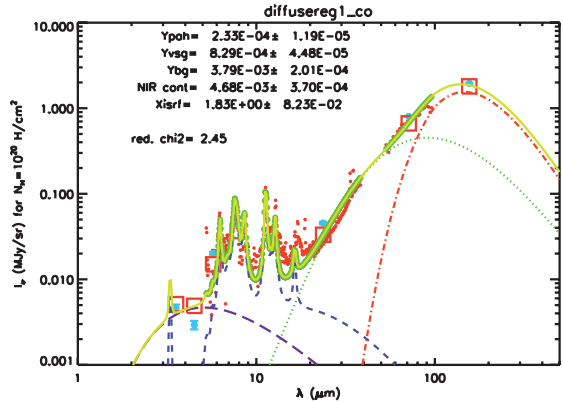}
\caption{Best fit derived using the \emph{DUSTEM} code for the
spectrum toward SSDR 1 (CO-154). The red dots show the IRS and MIPS
SED \emph{SAGE-Spec} spectrum integrated as discussed in the text. The
blue dots show the Spitzer photometry integrated in the SAGE data for
the same area. All values are normalized to a total column density of
$N_{\mathrm{H}} = 10^{20}$ cm$^{-2}$.  The various lines show the
contribution of individual dust components: red (dot-dash): Big
Grains, green (dashed): Very Small Grains, blue (dashed): PAH, violet
(long dash): NIR continuum. The total model spectrum is shown by the
continuous yellow line. The squares show the photometry in the Spitzer
bands derived from the model spectrum. The best fit values for the
free parameters shown are those of Table\,\ref{tab:DUSTEMFIT}.
\label{fig:DustemFit}}
\end{figure}

The integrated spectrum of \emph{SAGE-Spec} diffuse region \#1,
hereafter SSDR 1, which is also known as CO cloud 154 from the NANTEN
CO survey \citep{2008ApJS..178...56F}, or LMC N J0531-6830, with a
mass (estimated from the CO) of $1.0\times 10^6$ $M_\odot$
\citep{2008ApJS..178...56F}, is shown in Figure~\ref{fig:atomicspec}.
The spectrum is dominated by PAH features at 5--15 $\mu$m and a
continuum from small grains at longer wavelengths, while big grains
dominate the emission longward of $\sim$80 $\mu$m.  In order to
extract some physical properties of the PAHs and ionic lines in the
spectral cube, the spectrum was fitted using an adapted version of
\emph{PAHFIT} \citep{2007ApJ...656..770S} that handles both the IRS
(5--40 $\mu$m) and MIPS SED (70--90 $\mu$m) spectra.  To extract
abundances of the PAHs and grains populations responsible for the MIR
and FIR continuum emission, we used an empirical dust model,
\emph{DUSTEM} \citep[see][for a description]{2008AJ....136..919B}
which is based on the emission model of \citet{1990A&A...237..215D}.
We fit the integrated spectrum with the \emph{DUSTEM} model allowing
the radiation field intensity ($X_{\mathrm{ISRF}}$), the abundance of
the 3 dust components ($Y_{\mathrm{PAH}}$; Very Small Grains:
$Y_{\mathrm{VSG}}$; and Big Grains: $Y_{\mathrm{BG}}$) and the
intensity of the NIR continuum to vary simultaneously. The size
distribution of the PAHs and BGs remain in the form of a power law as
in the Solar Neighborhood \citep{1990A&A...237..215D}, but we model
the VSGs with a flatter size distribution to account for the flatter
shape of the far-infrared continuum: $n(a)da \propto a^{\alpha}$ with
$\alpha=-1$, where $a$ is the grain radius and $\alpha$ the index of
the size distribution of the VSGs.  The best fit is shown in
Fig.~\ref{fig:DustemFit} along with the \emph{SAGE-Spec} spectrum
normalized to a total dust column density of $10^{20}$ H cm$^{-2}$ as
derived from the H{\sc i} and CO maps. Fig.~\ref{fig:DustemFit} also
shows the photometry points derived from integrating the
\emph{SAGE-LMC} IRAC and MIPS maps over the same region used to
extract the spectrum.

\begin{deluxetable*}{lcccccc}
\tablenum{7}
\tablecaption{Physical parameters derived from the DUSTEM fit compared to those
obtained in \citet{2008AJ....136..919B} for cloud CO-154. The values labeled with '\ion{H}{1}' are run for a grain size distribution applicable to the Solar Neighborhood, while the values labeled 'CO' are those for \emph{case 2} as described by \citet{2008AJ....136..919B}, in which the same flattened grain size distribution is used for the VSGs as in the present work. \label{tab:DUSTEMFIT}}
\tablehead{\colhead{} & \colhead{$Y_{\mathrm{PAH}}$} & \colhead{$Y_{\mathrm{VSG}}$} & \colhead{$Y_{\mathrm{BG}}$} & \colhead{NIR cont} & \colhead{$X_{\mathrm{ISRF}}$} & \colhead{$\chi^2/\mathrm{dof}$}\\
\colhead{}& \colhead{$(10^{-4})$} & \colhead{$(10^{-4})$}
& \colhead{$(10^{-3})$} & \colhead{($10^{-4}$ MJy/sr)} & \colhead{} & \colhead{}}
\startdata
\emph{SAGE-Spec} & 2.33 & 8.29 & 3.79 & 46.8&
1.83 & 2.44 \\
CO-154(\ion{H}{1}) & 1.78 & 8.25 & 2.85 & 4.99 &
2.59 & 6.61\\
CO-154(CO) & 0.80 & 2.47 & 0.98 & 15.2 &
3.55 & 7.09 \\
\enddata
\end{deluxetable*}

The \emph{DUSTEM} fit results are given in Tab.~\ref{tab:DUSTEMFIT},
and shows that the dust abundances derived are somewhat in agreement
with the results given in \citet{2008AJ....136..919B} for cloud
CO-154, indicating that the region selected for spectral mapping is
representative of the larger region mapped in the \emph{SAGE-LMC}
imaging survey.  Note that the large value derived for the NIR
continuum probably reflects that the stellar contribution was not
removed from the data presented here. This does not affect the fit
parameters for the PAH, VSG and BG component, as the NIR wavelength
range is not used to derive the grain properties of those species.

\begin{deluxetable}{rcc}
\tablenum{8}
\tablecaption{Strengths of the major PAH features for SSDR 1, also known as NANTEN CO cloud
CO-154. \label{tab:PAHFIT}}
\tablehead{\colhead{Feature name} & \colhead{Strength} & \colhead{Uncertainty} \\ \colhead{} & \colhead{(W m$^{-2}$ sr$^{-1}$)} & \colhead{(W m$^{-2}$ sr$^{-1}$)}}\startdata
6.2 $\mu$m & 1290.8 & 1.2\\
7.7 $\mu$m Complex & 3374.8 & 5.2\\
8.6 $\mu$m & 1525.4 & 1.4\\
11.3 $\mu$m Complex & 2350.9 & 0.7\\
12.0 $\mu$m & 674.3 & 0.9\\
12.6 $\mu$m Complex & 945.9 & 0.9\\
13.6 $\mu$m & 453.6 &0.8\\
14.2 $\mu$m & 49.0 &.8\\
16.4 $\mu$m &171.5 &.4\\
17 $\mu$m Complex &668.7 &1.2\\
17.4 $\mu$m & 53.1 &0.4\\
\enddata
\end{deluxetable}

\begin{figure} 
\plotone{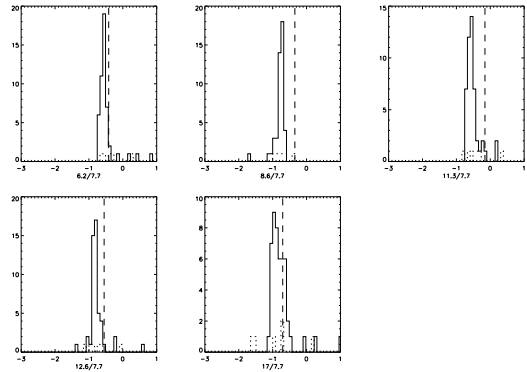}
\caption{Histograms of the ratios of PAH feature strengths from a
sample of nearby galaxies \citep{2007ApJ...656..770S}. Each ratio is
normalized to the 7.7 $\mu$m PAH complex.  The solid histograms are
for the 50 galaxies in the sample with [Ne III]$<$[Ne II], as observed
for SSDR 1, while the dotted histograms are for the 9 others.  The
feature ratios for SSDR 1 (CO-154) are indicated with vertical dashed
lines. \label{fig:diffPAHratio}}
\end{figure}

The intensities of the major PAH features derived using \emph{PAHFIT}
are shown in Table~\ref{tab:PAHFIT}.  Figure \ref{fig:diffPAHratio}
shows a comparison between the PAH strength ratios in SSDR 1 with
those of other nearby galaxies \citep{2007ApJ...656..770S}. The ISM
for all of these extragalactic spectra contain a range of radiation
fields. Although SSDR 1 is toward 
molecular cloud CO-154, there is clearly ionized gas in the line-of-sight,
based on the presence of the $[$\ion{Ne}{3}$]$ 15.5 and
$[$\ion{Ne}{2}$]$ 12.8 $\mu$m lines. The ratio of these lines,
$[$\ion{Ne}{3}$]$/$[$\ion{Ne}{2}$]$=0.17, indicating low-ionization
gas, is near the median of the ratios for the extragalactic
lines-of-sight studied by \citet{2007ApJ...656..770S}.  Nonetheless,
we separated the high-ionization lines-of-sight in order to keep the
comparison as even as possible. It is evident that the 8.6, 11.3 and
12.6 $\mu$m features are significantly stronger for SSDR 1 than the
average for the sample of nearby galaxies. The relatively bright 11.3
$\mu$m feature is characteristic of the ISM in the Magellanic
Clouds. The features were first detected outside \ion{H}{2} regions in
the SMC using ISO \citep{2000A&A...361..895R}, for a region similar in
many ways to SSDR 1 considered here; the feature ratios have been
interpreted as indicating a different intrinsic set of PAH band
strengths in the SMC as compared to those in the Milky Way
\citep{2002ApJ...576..762L}.

\subsubsection{Young Stellar Objects}

\begin{figure} 
\plotone{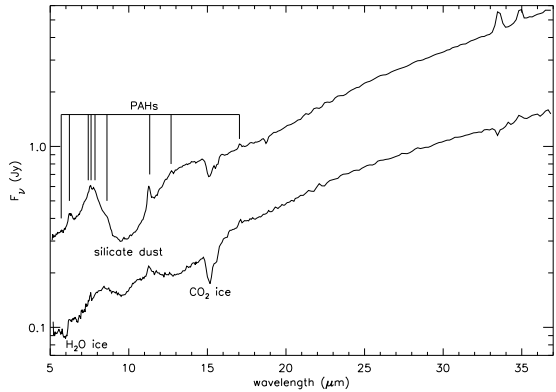}
\caption{Examples of IRS spectra for 2 embedded YSOs in the LMC
\citep{2009ApJ...707.1269O}.  Both objects exhibit a red dust
continuum characteristic of the early stages of YSO evolution.  Both
objects show the broad feature at 10\,$\mu$m associated with silicate
dust and PAH emission most noticeably at 6, 8 and 13\,$\mu$m. Their
spectra also show prominent ice signatures: CO$_2$ ice at
15.2\,$\mu$m, and the 5\,$-$\,7\,$\mu$m ice complex (for the bottom
object) that includes amongst others an important contribution from
water ice.
\label{fig:ysoice}}
\end{figure}

Fig.~\ref{fig:ysoice} shows two examples of embedded YSOs in the
\emph{SAGE-Spec} sample. Both spectra exhibit strong absorption at
15.2\,$\mu$m attributed to the CO$_2$ ice bending mode. These objects
also show strong silicate absorption at 10\,$\mu$m, as well as
emission features due to PAHs, more noticeable at 6, 8 and
11\,$\mu$m. The lower spectrum also seems to show an additional broad
absorption feature at 6\,$\mu$m. This ice band is mainly due to the
O$-$H bending mode of water ice, but it is also thought to include
contributions of other minor ice species.  We have analyzed in detail
the CO$_2$ profiles of a sample of massive embedded YSOs identified in
the \emph{SAGE-Spec} survey and from archival data, of which the
results are presented in another paper \citep{2009ApJ...707.1269O},
and summarized here.  We compute column densities and compare the
observed profiles with laboratory profiles available in the
literature. The observed profiles show a varied morphology that, when
modeled with the help of lab profiles, provides clues on the ice's
admixtures and environmental properties, like temperature. We also
investigate the properties of the 5$-$7 $\mu$m band. By comparing
observed properties in LMC and Galactic sample it is possible to get a
handle on metallicity effects on ice chemistry
\citep{2009ApJ...707.1269O}.

\subsection{The nature of point sources}
\label{sec:pointsources}

\subsubsection{MIPS SED point sources} Two examples of MIPS SED
spectra \citep{2010AJ....139...68V} are shown together with their IRS
spectra in Figure~\ref{fig:specps}. These two objects are very
different, both in their appearance and in their nature:
IRAS\,05280$-$6910 is an OH/IR star of high luminosity
\citep{1992ApJ...397..552W}; it is perhaps the most dust-enshrouded
supergiant known to exist in the Magellanic Clouds
\citep{2005A&A...442..597V}. We might be dealing with a flattened
circumstellar envelope, such as that of WOH\,G064
\citep{2008A&A...484..371O} but in this case viewed edge-on and thus
rendering the central star invisible at optical wavelengths. This
picture is confirmed by the large optical depth in the silicate
features in the IRS range and the absence of cold dust in the MIPS SED
range. The other object, IRAS\,05137$-$6914 is an ultra-compact
\ion{H}{2} region detected at radio wavelengths
\citep{1985ApJS...58..197M,2007MNRAS.378.1237B}. It must harbor a
massive young, hot star. The cold dust dominating the cocoon around
this young star emits an intense far-IR continuum. The excitation and
ionization conditions in the gas are traced by the fine-structure line
emission of [O{\sc i}] at 63 $\mu$m and [O{\sc iii}] at 88 $\mu$m in
the MIPS SED range; in the IRS range the [S{\sc iii}] lines at 18 and
33 $\mu$m are very bright and PAH emission as a consequence of the
erosion of small grains is seen. We present a thorough description of
the nature of {all} \emph{SAGE-Spec} MIPS SED point sources in a
related study \citep{2010AJ....139...68V}.

\subsubsection{A background quasar with peculiar properties}

\begin{figure}
\includegraphics[angle=90,scale=0.35]{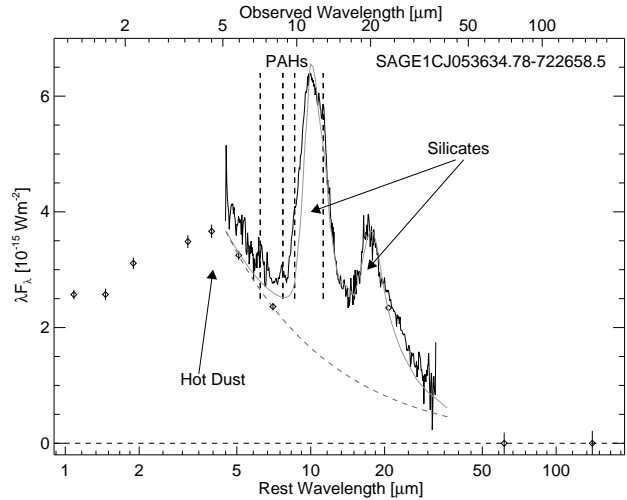}
\caption{Broad wavelength energy distribution of background quasar
SAGE 1C J053634.78$-$722658.5 combined with the remarkable IRS
spectrum of the source \citep{HKW_10_quasar}. The main contributions
that can be identified are the exceptionally strong silicate emission
bands at 9.7 and 18~$\mu$m and the hot continuum peaking near
4~$\mu$m. The weak PAH emission bands have been used to derive the
redshift of 0.14.
\label{fig:quasar}}
\end{figure}

The \emph{SAGE-Spec} program has also yielded IRS spectra of a number
of sources that could not easily be classified based on their
broad-band colors.  Several of these are background sources. One such
background source -- SAGE 1CJ053634.78$-$722658.5 -- stands out in the
IRS spectrum. It has a redshifted spectrum ($z$=0.14) that exhibits
extremely prominent silicate emission features at 10 and 18~$\mu$m. We
have analyzed the source and discussed its nature
\citep{HKW_10_quasar}. We argue that the peculiar IRS spectrum and its
corresponding broad wavelength energy distribution are indicative of a
quasar.  We do not detect any emission from the host galaxy; neither
the stellar component in the optical or near IR nor the colder ISM in
the far-IR \citep[see Fig.~\ref{fig:quasar};][]{HKW_10_quasar}, and
this may thus be another example of a host-less quasar
\citep{2005Natur.437..381M}.

\section{Outlook and conclusions}
\label{sec:conclusions}

The \emph{SAGE-Spec} program provides useful data for understanding
the life cycle of gas and dust in galaxies. The extensive dataset of
\emph{IRS} and \emph{MIPS SED} spectroscopy, obtained within the
\emph{SAGE-Spec} program, complemented with archival data, sample the
relevant environments and ultimately provides insights on dust
mineralogy and gas properties in these environments. Feeding the
results back to the original \emph{SAGE-LMC} data leads to conclusions
on stellar populations, and allow us to study the mineralogical dust
cycle in the Large Magellanic Cloud, in combination with the global
star formation rate
\citep[][]{2008AJ....136...18W,2009ApJS..184..172G} and injection rate
of stellar mass loss into the ISM
\citep[e.g.~][]{2009MNRAS.396..918M,2009AJ....137.4810S}.  An
important outcome of the \emph{SAGE-Spec} program is contributing
distinguishing diagnostics to classify sources in the \emph{SAGE-LMC}
point source catalog.

The initial results discussed in the paper include the first
extragalactic detection of the 21 $\mu$m feature; the study of
crystalline silicates in the disks around RV Tauri stars; the possible
detection of a host-less quasar; the analysis of ices towards massive
YSOs; and investigations into feature and line ratios in atomic and
\ion{H}{2} regions to probe physical conditions, such as radiation
field and ionization fraction.

True to its legacy status, the \emph{SAGE-Spec} team has delivered a
significant fraction of its reduced data to the scientific community
already, with further data deliveries planned in the near future. The
unprocessed data have been available to the community in the
\emph{Spitzer} archive from the date of observing. We will also
deliver enhanced data products, particularly spectral feature maps and
source and spectral classifications in those future deliveries.

\acknowledgements M.~Cohen thanks NASA for supporting his
participation in SAGE-Spec through JPL grant 1320707 with UC Berkeley.
B.~Sargent, M.~Meixner, and B.~Shiao were supported for SAGE-Spec
through JPL/SSC grant 1310534 with STScI.  M.~Meixner was additionally
supported by NASA NAG5-12595. R.~Szczerba acknowledges support from
grant N203 393334 (MNiSW).

\facility{Spitzer{IRS,MIPS}}



\clearpage
\LongTables
\begin{deluxetable}{@{}c@{}l@{}lllllll}
\tablenum{3}
\tabletypesize{\scriptsize}
\tablewidth{600pt}
\tablecaption{Point sources targeted with the IRS staring mode as part of the \emph{SAGE-Spec} program. \label{tab:irsss}}
\tablehead{\colhead{Nr.}&\colhead{AORkey} & \colhead{RA (J2000)} & \colhead{Dec (J2000)} &\colhead{SAGE designation} &\colhead{mod.} & \colhead{name} & \colhead{AOR key} & \colhead{Remarks}\\
&\colhead{IRS} &&& \colhead{SSTISAGE +} & & & \colhead{MIPS SED}}
\startdata
1&22399232&  04h37m21.15s& $-$70d34m44.57s &MC J043721.15$-$703444.7 &sl & NGC 1651 SAGE IRS 1&& Cluster                      \\
2&22399488&  04h37m27.69s& $-$67d54m34.94s &MC J043727.61$-$675435.1 &sl/ll & &22459648&               \\
3&22399744&  04h46m27.15s& $-$68d47m46.83s &MC J044627.10$-$684747.0 &sl/ll & &&        \\
4&22400256&  04h47m18.63s& $-$69d42m20.53s &MC J044718.63$-$694220.6 &sl/ll & &&          \\
5&24319488&  04h48m37.75s& $-$69d23m36.85s &MC J044837.77$-$692337.0 &sl/ll & &&       \\
6  &24318720&  04h49m34.38s& $-$69d05m49.17s &MC J044934.31$-$690549.3 &sl/ll & &&       \\
7  &22400512&  04h50m40.57s& $-$68d58m18.76s &MC J045040.52$-$685819.0 &sl/ll & MSX LMC 1128&&                            \\
8  &22400768&  04h51m28.58s& $-$69d55m49.92s &MC J045128.58$-$695550.1 &sl/ll & &&                                                 \\
\enddata
\end{deluxetable}

\LongTables
\begin{deluxetable}{ccccccccc}
\tablenum{4}
\tabletypesize{\scriptsize}
\tablecaption{Point sources targeted with the IRS staring mode as part of the \emph{SAGE-Spec} program, part II. \label{tab:irsss2}}
\tablehead{\colhead{Nr.}&\colhead{$J$} & \colhead{$H$} & \colhead{$K$} & \colhead{[3.6]} & \colhead{[4.5]} & \colhead{[5.8]}& \colhead{[8.0]}&\colhead{[24]}}
\startdata
1 & 12.41 & 11.49 & 11.19 & 10.91 & 10.81 & 10.69 & 10.49 & 9.64 \\
2 & 16.02 & 15.16 & 14.45 & 14.02 & 13.52 & 13.09 & 10.12 & 7.03 \\
3 & 15.02 & 13.28 & 11.72 & 9.86 & 8.97 & 8.55 & 8.12 & 7.33 \\
4 & 10.10 & 9.27 & 9.00 & 8.87 & 9.05 & 8.87 & 8.75 & 7.55 \\
5 & \nodata & \nodata & 12.92 & 10.36 & 9.48 & 8.68 & 7.85 & 4.86 \\
6 & 12.18 & 11.78 & 11.24 & 10.07 & 9.60 & 9.21 & 8.56 & 7.21 \\
7 & 13.22 & 11.56 & 10.37 & 9.37 & 8.94 & 8.57 & 8.14 & 7.62 \\
8 & 15.05 & 13.78 & 12.94 & 11.12 & 10.61 & 10.04 & 9.19 & 7.23 \\
\enddata
\tablenotetext{a}{MIPS-[24] from nearby LH$\alpha$ 120$-$N 82, which appears to be the main contributor to the LL spectrum.}
\end{deluxetable}

\LongTables
\begin{deluxetable}{rrrrlr}
\tablenum{5}
\tabletypesize{\scriptsize}
\tablecaption{Archival IRS staring mode observations within the \emph{SAGE-LMC} footprint \label{tab:archivalirs}}
\tablehead{\colhead{AOR key} & \colhead{RA (J2000)} & \colhead{Dec (J2000)} & \colhead{Proposal ID} & \colhead{PI} & \colhead{AOR key}\\ \colhead{IRS} & \colhead{requested} & \colhead{requested} & & & \colhead{MIPS SED}}
\startdata
        10972672&     04h28m30.17s& $-$69d30m50.0s  &            3426&          Kastner\\          
        19012352&     04h32m57.38s& $-$69d26m33.0s  &           30788&          Sahai\\          
        19012864&     04h33m43.69s& $-$70d09m50.0s  &           30788&          Sahai\\          
        19012608&     04h35m24.10s& $-$66d56m49.0s  &           30788&          Sahai\\          
        10972928&     04h37m22.73s& $-$68d25m03.3s  &            3426&          Kastner\\          
        24325120&     04h38m49.80s& $-$69d27m14.0s  &             464&          Cohen\\          
        23882240&     04h40m05.79s& $-$67d16m38.4s  &           40650&          Looney\\          
         6077440&     04h40m28.51s& $-$69d55m13.8s  &            1094&          Kemper\\
\enddata
\end{deluxetable}

\clearpage

\begin{deluxetable}{lllccccc}
\tablenum{6}
\tablecaption{Positions and map sizes of extended region observations in the \emph{SAGE-Spec} program. \label{tab:ext}}
\tablehead{\colhead{region} &\colhead{AOR key} & \colhead{AOR key} & \colhead{RA (J2000)} & \colhead{Dec (J2000)} & \colhead{size}\\
&\colhead{IRS}&\colhead{MIPS SED}
}
\startdata
SSDR\tablenotemark{a} 4&22461696 & 22465792   &04h47m40.85s& $-$67d12m31.0s & $1.0'\times 1.0'$ \\
DEM\tablenotemark{b} L 8& 22469120 & 22474240 &04h52m06.31s& $-$66d55m27.1s & 1.0$'\times 1.0'$ \\
DEM L 10& 22469376 & 22474496                 &04h52m11.90s& $-$69d20m43.1s & 2.3$'\times 1.0'$ \\
DEM L 34& 22469632 & 22474752                 &04h56m50.09s& $-$66d24m50.0s & 5.4$'\times 1.0'$ \\
DEM L 40& 22469888 & 22475008                 &04h57m41.21s& $-$67d38m53.9s & 2.1$'\times 1.0'$ \\
DEM L 55& 22470144 & 22475264                 &05h01m41.21s& $-$70d38m48.8s & 5.4$'\times 1.0'$ \\
DEM L 86& 22470400 & 22475520                 &05h09m55.99s& $-$68d54m03.2s & 4.6$'\times 1.0'$ \\
SSDR 3&22461440 & 22465536                    &05h15m43.64s& $-$68d03m20.3s & $1.0'\times 1.0'$ \\
\enddata
\tablenotetext{a}{\emph{SAGE-Spec} diffuse region}
\tablenotetext{b}{Nomenclature according to the catalog by \citet{1976MmRAS..81...89D}}
\tablenotetext{c}{This region is covered by a set of 14 IRS and 2 MIPS SED AORs, all part of Spitzer PID 30653 \citep{2009ApJ...694...84I}}
\end{deluxetable}

\end{document}